\input harvmac
\input epsf
\noblackbox
\def\IZ{\relax\ifmmode\mathchoice
{\hbox{\cmss Z\kern-.4em Z}}{\hbox{\cmss Z\kern-.4em Z}}
{\lower.9pt\hbox{\cmsss Z\kern-.4em Z}}
{\lower1.2pt\hbox{\cmsss Z\kern-.4em Z}}\else{\cmss Z\kern-.4em
Z}\fi}
\def\IB{\relax{\rm I\kern-.18em B}}
\def\IC{{\relax\hbox{$\inbar\kern-.3em{\rm C}$}}}
\def\ID{\relax{\rm I\kern-.18em D}}
\def\IE{\relax{\rm I\kern-.18em E}}
\def\IF{\relax{\rm I\kern-.18em F}}
\def\IG{\relax\hbox{$\inbar\kern-.3em{\rm G}$}}
\def\IGa{\relax\hbox{${\rm I}\kern-.18em\Gamma$}}
\def\IH{\relax{\rm I\kern-.18em H}}
\def\II{\relax{\rm I\kern-.18em I}}
\def\IK{\relax{\rm I\kern-.18em K}}
\def\IP{\relax{\rm I\kern-.18em P}}

\font\cmss=cmss10 \font\cmsss=cmss10 at 7pt
\def\IR{\relax{\rm I\kern-.18em R}}


\lref\ak{I.~Antoniadis and C.~Kounnas,
``Superstring phase transition at high temperature,''
Phys. Lett. {\bf B261}, 369 (1991).}

\lref\ang{C.~Angelantonj,
``Nontachyonic open descendants of the 0B string theory,''
Phys. Lett. {\bf B444}, 309 (1998)
hep-th/9810214}
\lref\sagss{
I. Antoniadis, E. Dudas,and  A. Sagnotti,
``Supersymmetry breaking, open strings and M-theory'',
Nucl.Phys. {\bf B544} (1999) 469-502, hep-th/9807011.}
\lref\sagssII{
I. Antoniadis, G. D'Appolonio, E. Dudas, and A. Sagnotti,
``Partial breaking of supersymmetry, open strings and M-theory'',
hep-th/9812118.}
\lref\bsag{M.~Bianchi and A.~Sagnotti,
``On The Systematics Of Open String Theories,''
Phys. Lett. {\bf B247}, 517 (1990).}
\lref\sag{A.~Sagnotti,
``Closed Strings And Their Open String Descendants,''
Phys. Rept. {\bf 184}, 167 (1989).}
\lref\adk{I. Antoniadis, J. Derendinger and C. Kounnas, ``Nonperturbative
Temperature Instabilities in N=4 Strings,'' Nucl. Phys. {\bf B551} (1999)
41, hep-th/9902032.}

\lref\oldpot{See for instance:
 V.A.~Kostelecky and S.~Samuel,
``The Tachyon Potential In String Theory,''
{\it Presented at 1988 Mtg. of Div. of Particle and Fields of the APS,
                  Storrs, CT, Aug 15-18, 1988}\semi 
T. Banks, ``The Tachyon Potential in String Theory,'' Nucl. Phys.
{\bf B361} (1991) 166\semi   
A. Belopolsky and B. Zwiebach, 
``Off Shell Closed String Amplitudes: Towards a Computation of
the Tachyon Potential,'' Nucl. Phys. {\bf B442} (1995) 494, hep-th/9409015.} 

\lref\pwit{J. Polchinski and E. Witten, ``Evidence for Heterotic-Type I
String Duality,'' Nucl. Phys. {\bf B460} (1996) 525, hep-th/9510169.}
\lref\bankssuss{T. Banks and L. Susskind, ``Brane - Anti-Brane Forces,''
hep-th/9511194.}
\lref\senbbbar{A. Sen, ``Tachyon Condensation on the Brane Anti-Brane
System,'' JHEP 9808 (1998) 012, hep-th/9805170.} 
\lref\yi{P. Yi, ``Membranes from Five-Branes and Fundamental
Strings from D-p Branes,'' hep-th/9901159.}
\lref\oldjoe{S.P. de Alwis, J. Polchinski, and R. Schimmrigk, 
``Heterotic Strings with Tree Level Cosmological Constant,''
Phys. Lett. {\bf B218} (1989) 449.}
\lref\hullrefs{C. Hull, ``The Nonperturbative SO(32) Heterotic String,''
hep-th/9812210\semi
E. Bergshoeff, E. Eyras, R. Halbersma, J.P. van der Schaar, 
C. Hull and Y. Lozano, ``Space-Time Filling Branes and Strings
with Sixteen Supercharges,'' hep-th/9812224.}
\lref\Osixteen{L. Alvarez-Gaume, P. Ginsparg, G. Moore and C. Vafa,
``An O(16) $\times$ O(16) Heterotic String,'' Phys. Lett. {\bf B171}
(1986) 155\semi
L. Dixon and J. Harvey, ``String Theories in Ten-Dimensions without
Space-Time Supersymmetry,'' Nucl. Phys. {\bf B274} (1986) 93.} 
\lref\aldazabal{G. Aldazabal, A. Font, L.E. Ibanez, A.M. Uranaga and
G. Violero, ``Non-Perturbative Heterotic D=6, D=4, N=1 Orbifold Vacua,''
Nucl. Phys. {\bf B519} (1998) 239, hep-th/9706158.}
\lref\kutasov{D. Kutasov, ``Orbifolds and Solitons,'' Phys. Lett.
{\bf B383} (1996) 48, hep-th/9512145.}
\lref\petr{P. Horava, ``Strings on World Sheet Orbifolds,'' Nucl. Phys.
{\bf B327} (1989) 461.}
\lref\joeetal{J. Dai, R.G. Leigh and J. Polchinski, ``New Connections
Between String Theories,'' Mod. Phys. Lett. {\bf A4} (1989) 2073.}
\lref\Kref{For a recent review with references, see: A. Sen, ``Non-BPS States and Branes
in String Theory,'' hep-th/9904207.}
\lref\joenotes{J. Polchinski, ``TASI Lectures on D-Branes,'' hep-th/9611050.} 
\lref\senorb{A. Sen, ``Duality and Orbifolds,'' Nucl. Phys. {\bf B474} (1996)
361, hep-th/9604070.} 
\lref\chs{C. Callan, J. Harvey and A. Strominger, ``World Sheet Approach to
Heterotic Instantons and Solitons,'' Nucl. Phys. {\bf B359} (1991) 611.}
\lref\paul{P. Aspinwall, ``Enhanced Gauge Symmetries and K3 Surfaces,''
Phys. Lett. {\bf B357} (1995) 329, hep-th/9507012.}
\lref\phases{E. Witten, ``Phases of N=2 Theories in Two Dimensions,''
Nucl. Phys. {\bf B403} (1993) 159, hep-th/9301042.}
\lref\berggab{O.Bergman and M. Gaberdiel, ``A Nonsupersymmetric Open
String Theory and S-Duality,'' Nucl. Phys. {\bf B499} (1997) 183, 
hep-th/9701137.}
\lref\lindil{R. Myers, ``New Dimensions for Old Strings,'' Phys. Lett.
{\bf 199B} (1987) 371\semi
I. Antoniadis,  C. Bachas, J. Ellis and D. Nanopoulos, ``Cosmological
String Theories and Discrete Inflation,'' Phys. Lett. {\bf 211B} (1988) 393.}
\lref\joeI{J. Polchinski, {\it String Theory}, Vol. I, Cambridge University
Press, 1998.}
\lref\kuthsu{D. Kutasov, ``Irreversibility of the Renormalization Group Flow
in Two-Dimensional Quantum Gravity,'' Mod. Phys. Lett. {\bf A7} (1992) 2943,
hep-th/9207064\semi
E. Hsu and D.
Kutasov, ``The Gravitational Sine-Gordon Model,'' Nucl. Phys. {\bf B396} (1993)
693, hep-th/9212023.}
\lref\nonsusy{See for instance: B. Kors, 
``D-brane Spectra of Nonsupersymmetric, Asymmetric Orbifolds and Nonperturbative
Contributions to the Cosmological Constant,'' hep-th/9907007\semi   
R. Blumenhagen and A. Kumar, 
``A Note on Orientifolds and Dualities of Type 0B String
Theory,'' hep-th/9906234\semi 
I. Klebanov, ``Tachyon Stabilization in the
AdS/CFT Correspondence,'' ~hep-th/9906220\semi
C. Angelatonj, I. Antoniadis and K. Forger, ``Nonsupersymmetric Type I
Strings with Zero Vacuum Energy,'' hep-th/9904092\semi 
R. Blumenhagen, A. Font and D. Lust, ``Tachyon Free Orientifolds of Type 0B
Strings in Various Dimensions,'' hep-th/9904069\semi
R. Blumenhagen and L. Gorlich, ``Orientifolds of Nonsupersymmetric 
Asymmetric Orbifolds,'' hep-th/9812158\semi
I. Antoniadis, G. D'Appolonio, E. Dudas, and A. Sagnotti,
``Partial breaking of supersymmetry, open strings and M-theory'',
hep-th/9812118\semi
I.R.~Klebanov and A.A.~Tseytlin,
``D-branes and Dual Gauge Theories in Type 0 Strings,''
Nucl. Phys. {\bf B546} (1999) 155,  
hep-th/9811035\semi
C.~Angelantonj,
``Nontachyonic open descendants of the 0B string theory,''
Phys. Lett. {\bf B444}, 309 (1998)
hep-th/9810214\semi
G. Shiu and S.H. Tye, ``Bose-Fermi Degeneracy and Duality in
Nonsupersymmetric Strings,'' Nucl. Phys. {\bf B542} (1999) 45, 
hep-th/9808095\semi 
S. Kachru and E. Silverstein, ``Self-Dual Nonsupersymmetric Type II String
Compactifications,'' JHEP 9811 (1998) 001, hep-th/9808056\semi
J. Harvey, ``String Duality and Nonsupersymmetric Strings,'' Phys. Rev.
{\bf D59} (1999) 026002, hep-th/9807213\semi 
M. Berkooz and S. Rey, ``Nonsupersymmetric Stable Vacua of M Theory,''
JHEP 9901 (1999) 014, hep-th/9807200\semi
S. Kachru, J. Kumar and E. Silverstein, ``Vacuum Energy Cancellation in a
Nonsupersymmetric String,'' 
Phys. Rev. {\bf D59} (1999) 106004, hep-th/9807076 \semi
I. Antoniadis, E. Dudas,and  A. Sagnotti,
``Supersymmetry breaking, open strings and M-theory'',
Nucl.Phys. {\bf B544} (1999) 469-502, hep-th/9807011\semi
S.~Kachru and E.~Silverstein,
``4-D Conformal Field Theories and Strings on Orbifolds,''
Phys. Rev. Lett. {\bf 80} (1998) 4855,  
hep-th/9802183\semi 
J.D.~Blum and K.R.~Dienes,
``Duality Without Supersymmetry: The Case of the $SO(16) \times SO(16)$
String,''
Phys. Lett. {\bf B414} (1997) 260,  
hep-th/9707148.
}

\Title{\vbox{\baselineskip12pt\hbox{hep-th/9907038}
\hbox{LBNL-43634, SLAC-PUB-8187, SU-ITP-99/34 }
}}
{\vbox{\centerline{Orientifolds, RG Flows, and Closed String Tachyons}
}
}                                   

\centerline{
Shamit Kachru,$^1$
Jason Kumar,$^2$
and Eva Silverstein$^3$
 }
\bigskip 
\centerline{$^{1}$Department of Physics}
\centerline{University of California at Berkeley}
\centerline{Berkeley, CA 94720}
\centerline{and}
\centerline{Ernest Orlando Lawrence Berkeley National Laboratory}
\centerline{Mail Stop 50A-5101, Berkeley, CA 94720}
\medskip
\centerline{$^{2}$ Department of Physics~~~~~$^3$ SLAC}
\centerline{~~~Stanford University}
\centerline{~~~Stanford, CA 94309}
\bigskip
\noindent

We discuss the fate of certain tachyonic closed string theories
from two perspectives.  In both cases our approach involves
studying directly  configurations
with finite negative tree-level cosmological constant.  
Closed string analogues of orientifolds,
which carry negative tension, are argued to represent the minima
of the tachyon potential in some cases.  In other cases, we
make use of the fact, noted in the early string theory literature, 
that strings can propagate  on spaces
of subcritical dimension at the expense of introducing
a tree-level cosmological constant.  
The form of the tachyon vertex operator in these cases makes
it clear that a subcritical-dimension theory results from tachyon
condensation.  Using results of Kutasov, we argue that
in some Scherk-Schwarz models, for finely-tuned tachyon condensates, 
a minimal model
CFT times a subcritical dimension theory results.     
In some instances,
these two sets of ideas may be related by duality.

\Date{July 1999}

\newsec{Introduction}

Many closed string theories contain tachyons.
In addition to the bosonic string, numerous orbifold models
have tachyons appearing in a twisted sector.
Although there are also many string backgrounds
which do not contain tachyons in the perturbative spectrum,
it is potentially of interest to understand the fate of those which do.  

Some work that was done earlier involved
attempting to study the tachyon potential directly through (off-shell)
calculations of the tachyon effective potential \oldpot.  In this
paper we take a somewhat different approach to this   
question (using a combination of orientifold physics and the observations
in \oldjoe) and discuss some examples which
realize it.  
A basic issue that arises
is the following.  A perturbative closed string
compactification has vanishing tree-level vacuum energy (formally all
one-point functions, including those of the dilaton and graviton,
vanish in the closed-string sphere diagram).  Therefore the
tachyon potential, at fixed dilaton vacuum expectation 
value, looks like figure 1 in the regime accessible to
the perturbative closed string description; in particular
the tachyonic maximum of the potential is at $V(0)=0$.  Then tachyon
condensation leads to configurations with negative energy.
Our main question is whether (1) there is a minimum configuration 
with finite negative energy into which the system can roll (and then
what the eventual solution of the dilaton equation of motion
is) or whether 
(2) the tachyon rolls to
negative infinity.

\smallskip
\epsfbox{fig1.eps}
\smallskip
\centerline{Figure 1: Tachyon on top of potential hill}
\smallskip

We will here consider two different ways of obtaining 
configurations with finite negative tree-level cosmological constant.
The first involves the physics of orientifolds (and analogous
configurations in closed string theories); the second involves
strings propagating on spaces of subcritical dimension.
As we will mention below, these two methods may in fact be
related to each other by dualities.

In the context of open string theories orientifold backgrounds
\refs{\sag,\petr,\joeetal}  have
the feature one would need for possibility (1), namely
a negative contribution to the tree-level vacuum
energy, which is finite for fixed nonzero string coupling.  
In \S2 we will set up an open string theory
which involves orientifolds, anti-orientifolds, branes,
and anti-branes in which the potential looks like
figure 1.  This model is similar to one discussed
earlier in the context of string-string duality
by Bergman and Gaberdiel \berggab\ (and formulated 
earlier by Bianchi and Sagnotti \bsag).  Then using locally the mechanism of 
\refs{\bankssuss,\senbbbar,\yi}, we will see that the
tachyons condense, leaving the system in a finite-negative-energy
minimum involving only orientifolds and anti-orientifolds.  

We find that in some closed string theories, 
non-perturbative configurations, analogous
to orientifolds in the open string case, exist and
can play a similar role.  These configurations include
the S-duals of orientifold planes \refs{\kutasov,\hullrefs}
and non-level-matched but anomaly free orbifold backgrounds
such as those of \aldazabal.  We will discuss these
configurations, which look locally like
figure 2, in \S3.  In this approach, problem is then to relate 
configurations of the type in figure 1 with those
of the type in figure 2.  

\smallskip
\epsfbox{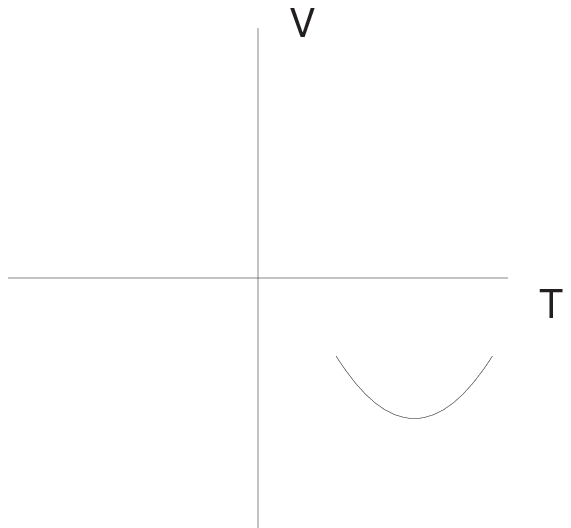}
\smallskip
\centerline{Figure 2: Tachyon at minimum of its potential}
\smallskip

The simplest such examples are given by the S-dual
of the open string theory of \S2\ (and various heterotic cousins
of this S-dual model).
We discuss these theories in \S4.
They involve orbifold fixed points whose degrees of
freedom microscopically represent the collective coordinates
of NS five-branes sitting at the singularities \kutasov, as well as
other orbifold fixed points, breaking a complementary
half of the supersymmetries from the first set,  
where NS anti-five-branes sit.  Globally this system is 
unstable, and can undergo NS brane-anti-brane annihilation.  
Condensation of the unstable mode therefore
leaves a vacuum with S-dual orientifolds and anti-orientifolds; 
the negative (S-dual) orientifold
tensions account for the negative vacuum energy after tachyon condensation. 

Our second approach, based on \oldjoe, is the following.
An a priori different way of getting backgrounds with finite negative
tree-level cosmological constant was explained in \oldjoe, in
which corresponding solutions to the dilaton and graviton
equations of motion (linear dilaton solutions) were obtained.  
(The latter step in particular eliminates the dilaton and graviton
tadpoles which we ignored in the preceeding discussion.)
These solutions involve strings propagating on spaces of sub-critical
dimension, in which the worldsheet beta functions are cancelled
by the contributions of dilaton gradients.  In \S5 we will
argue that a large class of tachyonic theories can be seen to
flow to such backgrounds upon tachyon condensation.  This analysis
uses simple features of the closed string tachyon vertex operators
in this class of examples. In particular generically these vertex operators
constitute (or generate) mass terms lifting some of the degrees
of freedom on the worldsheet, and one is left after renormalization-group
flow with a theory of sub-critical dimension of the type considered
in \oldjoe. For instance, we will see in \S5\ that 
by tuning the tachyon condensate in
some Scherk-Schwarz models one
can obtain minimal model CFTs times a sigma model on a subcritical
dimension target space.  A connection between non-critical
string theory and tachyon condensation (in the context of
the Hagedorn transition) was earlier conjectured in \ak,
where the form of the tachyon potential in heterotic Scherk-Schwarz
compactifications was explored.  
We further find that this point of view is potentially related
to the above point of view involving orientifolds if the
type of duality conjectured in \berggab\ is realized in string
theory.

Before proceeding we should remark on the dilaton direction in
the potential.  Any Poincare-invariant configuration
with negative tree-level vacuum energy is not a solution to
the equations of motion.  In string frame the vacuum energy is
\eqn\dilpot{
\Lambda={1\over g_s^2}\Lambda_0+\dots
}
So in order to solve the equations of motion the
dilaton $\phi$ must vary over spacetime.  
Similar remarks apply to the spacetime metric $G_{\mu\nu}$.
As we will discuss in \S5, the full solution
may be obtained through the conformal field theory techniques
of \oldjoe.  However at some points in our analysis, particularly
involving our first approach to the problem (in \S2-\S4), we
will attempt to separate the dependence of
the spacetime potential on
the originally tachyonic mode from its dependence on the dilaton and
metric by 
holding $\phi$ and $G_{\mu\nu} = \eta_{\mu\nu}$ fixed 
artificially.  The true
dynamics of the system will of course involve all of these fields,
as discussed more thoroughly in \S5.   

Many interesting papers which discuss tachyons on branes
in open string theories (``field theoretic'' tachyons) have
recently appeared; see e.g.
\Kref.  There have also been interesting recent papers
on other types of non-supersymmetric string backgrounds \nonsusy.

\newsec{An illustrative open-string theory}

Consider the type IIB string theory compactified on $T^d/I_d\Omega$,
where $I_d$ is a reflection on $d$ coordinates and $\Omega$ is 
a reversal of orientation on the IIB worldsheet.  The fixed planes
of the $I_d\Omega$ action, the orientifold $(9-d)$-planes,
have effectively negative tension \joenotes; that is, their
graviton and dilaton one-point functions have the opposite sign
to those of $D(9-d)$-branes.  There being no degrees of freedom
tied to the orientifold planes, this leads to a negative contribution
to the tree-level vacuum energy without leading to any unphysical
negative kinetic energy.  These orientifold planes are also charged
oppositely to the $D(9-d)$-branes under the $(10-d)$-form RR potential.
Therefore in the compact $T^d$, in order for RR flux lines to
end consistently, we must introduce 16
$D(9-d)$-branes.  In addition to cancelling
the RR charge, the positive tensions of the branes
cancel the negative tree-level vacuum energy of the orientifolds.
This example is very well known, as a T-dual of
the type I string theory on $T^d$.

Now consider instead a $Z_2\times Z_2$ orientifold of type IIB
(studied in \sagss).  
on $T^d$ generated by 
\eqn\gone{g_1=I_d\Omega}
\eqn\gtwo{g_2=I_d\Omega (-1)^{F}\delta}     
where $\delta$ is a translation by halfway around one of
the circles, with say coordinate $x_1$, 
of the $T^d$.  The element $g_2$ preserves
the opposite half of the supersymmetry from the half
preserved by $g_1$; it introduces {\it anti}-orientifolds
at positions halfway around the $x_1$ circle from the
positions of the orientifold planes introduced by $g_1$. 
The element $g_2$ alone would require the introduction of
16 {\it anti}-D$(9-d)$ branes.  

Given both the orientifolds and the anti-orientifolds, we
now have many options.  We could: 

A) Add the 16 D$(9-d)$-branes and 16 anti-D$(9-d)$-branes
corresponding to each orientifold action alone, and
then project the resulting open string spectrum onto
invariant states.  This
leads to a theory with RR charge conserved, zero tree-level
vacuum energy, and gauge group $O(16)\times O(16)$ or a rank-16
subgroup depending on the positions of the branes.  There is also
a 
tachyon multiplet in the ${\bf (16,16)}$ of $O(16)\times O(16)$,
for sufficiently small radius or sufficiently small separation 
of branes and anti-branes, and a closed string tachyon for
sufficiently small radius.    

Alternatively we could: 

B)  Add $k<16$ D$(9-d)$-branes and $k$ anti-D$(9-d)$-branes.
Again the RR charge is conserved, but the tree-level vacuum
energy is negative (as there are too few branes to cancel
the negative tension from the orientifolds).  The gauge
group is now $O(k)\times O(k)$ and there is a
scalar in the bifundamental which becomes tachyonic in
appropriate regimes of the classical moduli space where the
branes and anti-branes are close enough to each other.

It is clear in this example that we can annihilate $16-k$ of
the branes against $16-k$ of the antibranes, which proceeds
via condensation of $(16-k)\times (16-k)$ components of the bifundamental
tachyon along with condensation of the appropriate magnetic tachyon \yi.
In particular when all the open string tachyons of A) condense,
we are left with only orientifolds and antiorientifolds,
no gauge group, and finite negative vacuum energy (for
fixed nonzero dilaton).  So in this open-string context, the
zero-energy tachyonic theory A) rolls to a finite negative
energy configuration B) (plus a gas of excitations, since 
energy is conserved in this process).

\newsec{Negative-tension defects in closed string theory}

Now, we turn to a discussion of closed string theories with tachyons. 
It is natural to ask if there are negative-tension
configurations in closed string theory, analogous to the orientifolds of \S2,
to which such theories could roll.  

\subsec{S-duality and Orientifolds}

One obvious place to look for such a configuration is from
the S-dual of an orientifold plane in type IIB theory \refs{\kutasov,\senorb}.
Consider for example an orientifold 5-plane.  In the  
IIB theory on $\IR^4/I_4\Omega$, cancelling the RR charge
locally gives a D5-brane at an orientifold 5-plane.
The S-dual of this orientifold is type IIB on the orbifold 
$\IR^4/I_4(-1)^{F_L}$.  The twisted sector of this orbifold
constitutes precisely the worldvolume theory on an NS 5-brane at
a $Z_2$  singularity.  Giving vacuum expectation values
to the twisted sector scalars corresponds to moving the NS 5-brane
off of the fixed point.  The $Z_2$ fixed point which is left
(which has no degrees of freedom associated with it perturbatively --
all of the ``twisted sector moduli'' are accounted for by the
translation modes of the NS 5-brane)
is the S-dual of the orientifold.  The full closed-string orbifold
theory has zero vacuum energy,  so since the NS 5-branes have positive
tension (measured by an integral on a sphere surrounding the
brane), the S-dual of
the orientifold contributes negatively to the vacuum energy.
So one expects, for instance, that a dilaton gradient should emerge as one
turns on the marginal perturbations in the twisted sector of
the orbifold (moving the NS brane away from the S-dual orientifold),
in analogy with \pwit.  It would be very interesting to understand
how this works out in detail.     


\subsec{More general examples from non-perturbative orbifolds}

The S-dual orientifold is an example of a larger class of
negative-tension backgrounds.  In perturbative string theory,
modular invariance imposes conditions which sometimes go
beyond the physical condition of anomaly cancellation.
Modular invariance in the context of orbifold models requires
the inclusion of twisted sectors.  In the S-dual orientifold
example, after removing the NS 5-branes we are left with an
orbifold fixed point with no twisted states.  As we discussed
above this object exists in the theory and has a negative
contribution to $\Lambda_{tree}$.  

More generally we can consider orbifold models in which the
level-matching conditions, which ensure that physical perturbative
string states exist in the twisted sectors, are violated \aldazabal.
For example, let us consider the heterotic theory compactified
on a K3 surface realized as an orbifold $T^4/\IZ_{2}$ (our comments 
generalize immediately to the $T^4/\IZ_{k}$ orbifold realizations
of K3 with $k\neq 2$ as well).
Green-Schwarz anomaly cancellation requires only that
\eqn\gs{dH=tr R\wedge R-tr F\wedge F-n_5\delta_5}
where $\delta_5$ denotes delta functions localized at the positions
of the $n_5$ fivebranes. 

On the compact K3, $\int dH=0$, so $\int tr F\wedge F +n_5
= \int tr R\wedge R$.
This means that the number of instantons in the gauge bundle,
$n_{inst}$, and the number of 5-branes $n_5$ are constrained by  
\eqn\gsint{n_{inst}+n_5=24.}    
The perturbative string orbifold level-matching conditions
require the orbifold group to act nontrivially, according
to one of a few discrete possibilities, on the gauge degrees
of freedom of the heterotic string.  This
enforces the introduction of 24 instantons.  Non-perturbative
orbifold backgrounds with fewer instantons and a compensating
number of 5-branes involve non-modular-invariant
choices of the action on the gauge degrees of freedom.  
These backgrounds fit into an intricate web of six-dimensional
string dualities \aldazabal.  Again in these cases, the fact that the
5-branes have positive tension, and therefore contribute positively
to $\Lambda_{tree}$ in these backgrounds, means that the non-level-matched
orbifold fixed points that are left over contribute negatively
to $\Lambda_{tree}$.  

In fact, even without invoking non-perturbative physics 
we can find configurations with
effectively negative tension.  Suppose for example that we start
with an orbifold action with the standard embedding in the gauge
degrees of freedom of the heterotic string.  In this case, $dH$
cancels locally.  Now consider moving some large instantons off
the orbifold fixed points.  These instantons constitute gauge
fivebranes \chs\ of the heterotic string, and have positive
tension.  Since the overall vacuum energy of the string vacuum
is zero, the contribution of the remaining orbifold fixed points
is negative.  Clearly there is a whole zoo of possibilities of
this sort.  Furthermore as long as the instantons are large,
the whole configuration is accessible perturbatively.\foot{We
thank P. Aspinwall and R. Plesser for discussions on these points.}  

\subsec{S-dual O9-planes?}

Hull and collaborators \hullrefs\ have proposed formulating
the $SO(32)$ heterotic string as an orbifold of
type IIB by $(-1)^{F_L}$.  In perturbative
string theory, where we must add the twisted sectors dictated
by modular invariance, orbifolding by $(-1)^{F_L}$ produces IIA
from IIB and vice versa.  

On the other hand, there are other solutions to the anomaly
cancellation conditions.  In particular, one could add 
an $SO(32)$ or $E_8 \times E_8$ gauge group, obtaining the heterotic
string theory.  In the $SO(32)$ case, it is argued in \hullrefs\  
that this background can be deconstructed into an
S-dual O9-plane plus (in our way of counting) 16 NS 9-branes.  
In particular,
by S-dualizing the Born-Infeld like action 
for the 
ordinary O9-plane plus 16 D9-branes, one obtains this formulation
of the heterotic string.
In this point of view, the S-dual O-9-plane has negative tension,
$\Lambda_{S-O9}<0$, which is cancelled
by the tensions of the 16 NS 9-branes to give a vacuum with
$\Lambda_{tree}=0$.  These branes, if they exist, may provide
further possibilities for the endpoint of tachyon condensation
in various backgrounds.

\newsec{Examples of tachyon condensation}

In the last section we saw that there exists 
a plethora of possible configurations of negative 
vacuum energy (a la figure 2), realized microscopically
as S-duals of orientifolds and their generalizations.  
In this section we consider the problem,
on which we will only make a start here,
of understanding which tachyonic models (fig 1) roll
to to which (if any) of these negative-energy configurations.
In \S5\ we will consider another type of negative-energy
configuration which can plausibly be the endpoint
of tachyon condensation for a different class of theories.
We have mostly been confined here to
an analysis of orbifold backgrounds. One can hope that once
more generic non-supersymmetric compactification geometries 
are understood, there will be more general examples of such backgrounds.
Similarly many of the tachyonic theories (fig 1) that we are interested
in explaining are perturbative string orbifolds.  In general,
as twisted sector tachyon condensation will break the quantum
symmetry of the orbifold, we would expect orbifold models
to roll to more generic geometrical backgrounds.
  
So we will here consider a non-supersymmetric theory whose tachyons
arise at a more generic point on the CFT moduli point (away from
an orbifold point).  In particular, let us consider the non-supersymmetric
model which is S-dual to the $d=4$ ($d$ is the number of
compact dimensions) case of the orientifold model
of \S2.  We begin with type IIB theory on $T^4$ and
mod out by a $\IZ_2\times \IZ_2$ symmetry generated by 
\eqn\orbex{g_1 ~=~ I_4(-1)^{F_L},~~g_2 ~=~ I_4 (-1)^{F_R}\delta_1.}
Here $\delta_1$ is say a shift halfway around one of the circles of
the torus.
This theory has 16 orbifold
fixed points (8 fixed points each for the $g_{1}$ and $g_{2}$ actions). 
Based on the analysis of \kutasov, we find that
there are NS fivebranes sitting at the $g_1$ fixed points and 
NS anti-five-branes sitting at the $g_2$ fixed points. 
The vacuum energy is zero, and for large enough radius there is
no tachyon in the $(-1)^F\delta_1$ twisted sector.  
At genus zero, this theory is consistently described by 
nonsingular orbifold
conformal field theory.    

Globally the set of fivebranes and anti-five-branes is topologically
indistinguishable from the vacuum.  This being true at the level of
the classical field configuration, we expect to find an instability
in this system at some order in conformal perturbation theory
about the orbifold limit.  The leading possibility is that moving
some combination of fivebranes off the fixed point(s)--an exactly
marginal perturbation--induces a
tachyonic mass term for the collective coordinate describing the motion
of the anti-five-branes off the orbifold points.  This 
corresponds to a force on the anti-five-branes linear in
their displacement from the orbifold points.  
There is such a force in some regimes at the level of the low-energy
gravity theory, valid when all distances are much
larger than string scale.  There is globally therefore
a manifest instability
in the system, though $\alpha^\prime$ corrections may come
in at substringy distances and lift the tachyonic mode
from the point where the anti-fivebranes are at the orbifold points.  
In any case it is natural to postulate that the fivebranes and anti-fivebranes
annihilate, leaving behind the S-dual orientifolds which are hidden 
at the orbifold fixed points.

Given this, this example gives us one case where the result
of condensation of a negative mode, most probably a tachyon, 
is a configuration of
finite negative tension of the type discussed in \S3\
(at fixed dilaton) .  
Many other similar models can be obtained by considering the heterotic
theory on $T^4$ modded out by $I_4$ and $I_4 (-1)^{F}$, with
a level-matching action on the gauge bundle that introduces
24 instantons and 24 anti-instantons.  These then can move off
the fixed points and enlarge; there is a global instability
in the moduli space which we would expect to see perturbatively.
The global minimum of the potential for the non-dilatonic moduli
is again a configuration of finite negative tension of the
type discussed in \S3.  We leave detailed analysis of this class
of examples for future work.


\newsec{Tachyons and RG Flow}

Another context in which backgrounds with finite negative tree-level
cosmological constant arise is the following \oldjoe.  Consider
bosonic strings
propagating in a spacetime of dimension $D<D_{crit}$.\foot{Our
discussion applies with minor modifications to other strings as well.} 
The
Weyl anomaly cancellation conditions, which include
\eqn\dilbeta{
0\equiv\beta^{\Phi}={{D-D_{crit}}\over 6}-
{\alpha^\prime\over 2}\nabla^2\Phi+\alpha^\prime\nabla_\mu\Phi\nabla^\mu\Phi
-{\alpha^\prime\over 24}H^2+O({\alpha^\prime}^2)}
can still be solved in such a situation by allowing the
dilaton to vary over spacetime.  For example one exact CFT
solution is the linear dilaton background \refs{\oldjoe,\lindil}.    
The Weyl anomaly conditions, regarded as equations of motion
for the spacetime theory, follow from a Lagrangian
\eqn\spaceact{
S\propto
\int d^Dx \sqrt{-G} e^{-2\Phi}
\biggl(-{{2(D-D_{crit})}\over{3\alpha^\prime}}+R-{1\over {12}}H^2
+4\partial_\mu\Phi\partial^\mu\Phi+O(\alpha^\prime)\biggr)
 }
The first term is a finite negative cosmological constant
proportional to $D-D_{crit}$.  

It was noted in \oldjoe\ that such sub-critical-dimension theories
could arise naturally as the result of tachyon condensation.
Tachyon vertex operators are relevant operators of the internal
worldsheet CFT (that is the part of the worldsheet CFT not
involving the noncompact Poincare-invariant spacetime).  

Let us consider a specific case.  Take the bosonic string theory
compactified on the $SO(32)$ lattice.  In a fermionic description
the $SO(32)^2$ current algebra arises from 32 real left-moving
fermions $\lambda^I, I=1,\dots,32$ and 32 real right-moving fermions
$\tilde\lambda^{\tilde I}, {\tilde I}=1,\dots, 32$.  
There are in addition 10 
scalar fields $X^\mu, \mu=0,\dots,10$  
making up the rest of the 26 units of central charge.
This theory has in addition to the 
singlet tachyon (the universal tachyon of the
bosonic string) a ${\bf (32,32)}$ tachyon.  The vertex operator
for the latter is 
\eqn\bosvert{
V_{(32,32)}=\lambda^I\tilde\lambda^{\tilde J} e^{ik\cdot X}
}

To describe tachyon condensation, at leading order we wish to
add the integrated tachyon vertex operator to the worldsheet
action.  At zero momentum $k$ this operator is relevant, and
the covariant vertex operator which we would naively add,
\eqn\vertcov{
\int d^2\sigma \sqrt{g} \lambda\tilde\lambda}
has nontrivial dependence on the conformal factor $\omega$ in
the worldsheet metric.  (Here we use diffeomorphisms
to fix $g=\eta e^{2\omega}$ where $\eta$ is the flat metric
on the worldsheet.)  Heuristically, since \vertcov\ is
a mass term for $\lambda$ and $\tilde\lambda$, we 
expect this relevant deformation to lift these degrees
of freedom from the worldsheet theory, giving us a string
theory propagating effectively in $D=26-{N\over 2}$ where
$N$ is the number of worldsheet fermions that pair up and
become massive.  

However since this procedure involves going off shell,
as formulated here it breaks the Weyl symmetry of
the worldsheet theory classically.  When we have a well-defined
on-shell string theory, before condensing the tachyon, 
the formulation of the worldsheet theory involves
dividing the worldsheet path integral out by the volume
of the Weyl group, thereby eliminating the conformal 
factor $\omega$ from the theory.  It is difficult
to see how this degree of freedom could be introduced in
a physically continuous deformation from that point.

Let us therefore consider adding instead\foot{We thank
S. Shenker for a discussion on this approach.}
\eqn\tachfull{
\int d^2 \sigma\sqrt{g} \lambda\tilde\lambda e^{ik\cdot X}
}
with $k^2=m^2<0$.  This operator is dimension (1,1) and
the expression is Weyl invariant (due to the metric-dependence
in the regulated operators \joeI ).  In particular let us 
take $k=(\pm i\kappa,0,\dots,0)$ with $\kappa$ positive.  
This gives us a dimension
(1,1) operator describing a time-dependent
tachyon condensate.  Then the terms we can add are
of the form
\eqn\timeterm{
\int d^2 \sigma\sqrt{g} \lambda\tilde\lambda (a_+ e^{\kappa X_0}
+a_-e^{-\kappa X_0})
}    
for some parameters $a_{\pm}$.  

The addition \timeterm\ describes the initial time-dependence of
the tachyon as it begins to roll down its potential hill.
Let us take $a_- ~=~0$ so that the condensate vanishes 
at very early times $X_0 \to -\infty$.
Though as the condensation proceeds we expect higher order corrections
to \timeterm, we do expect that the tachyon will continue to
condense (rather than returning to its unstable maximum) and
the coefficient of the $\lambda \tilde \lambda$ mass term in
\timeterm\ will be nonzero for $X_0 > -\infty$.
Therefore it seems quite
plausible that the effect of tachyon condensation is to lift
the degrees of freedom $\lambda, \tilde\lambda$ and correspondingly
introduce a nonzero $D-26$ term in the beta function equations
(spacetime equations of motion).  In order to preserve conformal
invariance, therefore, the spacetime dilaton needs to begin
varying so as to maintain $\beta^\Phi=0$.  (In other words
we are adding a marginal perturbation \tachfull\ and therefore
must maintain that the total central charge is constant.)
One example of a solution to the modified equations of motion would,
for instance, be the linear dilaton solution of \refs{\oldjoe,\lindil},
though it is not clear that the dynamics favors this solution.

\subsec{Relation to orientifold picture?}

This lifting of the $\lambda,\tilde\lambda$ degrees of freedom
agrees with the picture one would obtain from the relation of
this theory to a certain type I orientifold conjectured by 
Bergman and Gaberdiel 
\berggab.  The orientifold in question is simply an orientifold
of type IIB by a $\IZ_2\times\IZ_2$ action generated by 
$(-1)^F$ and $\Omega$.  This theory (which is extremely similar 
to the one we discussed in \S4, differing by a shift and some duality
transformations)
has an orientifold 9-plane, an anti-orientifold
9-plane, 16 D9-branes and 16 anti-D9-branes.  This gives a
gauge group $SO(32)^2$ with tachyons in the ${\bf (32,32)}$ representation.
This theory contains in its spectrum of D1-branes one which has the
worldvolume degrees of freedom of the bosonic string compactified
on the $SO(32)$ lattice.  These degrees of freedom consist
of 10 bosonic collective coordinates from the 1-1 sector of
open strings, and 32 left-moving fermions from the 1-9 sector
and 32 right-moving fermions from the $1-\bar 9$ sector.  

In this theory it is relatively clear what happens when the ${\bf (32,32)}$
tachyon condenses:  the D9-branes annihilate with the anti-D9-branes.
This leaves the spacetime theory with the O9 and anti-O9 planes,
which contribute a finite negative amount to the tree-level
cosmological constant as discussed in very similar circumstances above.

Now let us consider what happens on the D1-brane probe in the orientifold
theory.  Once the D9-branes and anti-D9-branes have annihilated,
the contributions from the 1-9 and $1-\bar 9$ sectors disappear.
One is left with a low-energy worldvolume theory with 10 scalar
collective coordinates $X^\mu, \mu=1,\dots,10$.  This gives
the same result as we found for the bosonic string in the
RG flow analysis (in particular the absence of
both the fermions $\lambda$ and $\tilde\lambda$ and the decoupling of
the Liouville mode classically on the worldsheet).  
Therefore if the duality between the two theories conjectured
in \berggab\ turns out to hold\foot{The meaning of this duality
proposal is not entirely clear, since the dilaton potential is not flat in
the proposed duals.} 
there is a relation between
the two approaches we are considering in this paper.

\subsec{Other Models}

There is a large set of models amenable to the analysis
involving RG flow.  For example, let us consider a (T-dual version
of a) Scherk-Schwarz
compactification.  This consists of an orbifold of the
toroidally compactified type II or heterotic theory by $(-1)^F\delta$
where $\delta$ is a level-matched shift symmetry of the Narain
lattice.  
We choose $\delta$ so that the Scherk-Schwarz tachyon(s) are
momentum modes 
which becomes tachyonic at sufficiently large radius.
The integrated vertex operators for these tachyons take the form
(in the type II case for specificity)
\eqn\ssvert{
\int d^2zd^2\theta \sum_p\lambda_p cos{[(p+\delta^\prime+k)\cdot {\bf X}]}
+\rho_p sin{[(p+\delta^\prime+k)\cdot {\bf X}]}}  
Here we take $\delta^\prime$ to be half a momentum lattice vector, 
and we sum over those $p$ in the momentum lattice such that the internal
piece of the added operator 
is relevant--i.e. we only include tachyons.  $k$ is
spacetime momentum in the noncompact directions. We work in
$(1,1)$ superspace with 
${\bf X}=X+\theta\psi+\bar\theta\tilde\psi+F$ a scalar superfield.  
In the $(0,0)$ picture, in components,  the vertex operator is
\eqn\zerovert{
V^{(0,0)}=(\delta^\prime_L+k)\cdot \psi (\delta^\prime_R+k)\cdot\tilde\psi
e^{i\delta^\prime\cdot X} e^{ik\cdot X}
}
The renormalization group flow for this type of potential has been studied
by Kutasov \kuthsu.  He finds generically a mass gap for the degrees
of freedom in ${\bf X}\cdot\delta^\prime$.  With some fine-tuning
of the coefficients $\lambda$ and $\rho$, 
one can obtain minima in the potential which locally take
the form of ${\bf X}^n$ potentials for various $n>2$.  Thus
minimal models arise as the result of tachyon
condensation in these special directions.  In such cases,
although one does not lift all the central charge of
the ${\bf X}$ multiplet, it decreases in the flow 
to the value of the appropriate minimal model, and
the orthogonal spacetime part of the theory is still of sub-critical
dimension.   

A similar analysis can be done for heterotic compactifications
of the form $T^d/(-1)^F\delta$.     
More generally, this set of ideas might be helpful in
any situation where the closed string tachyon is charged under some
spacetime gauge symmetry.  
Such theories with charged tachyons 
are natural candidates for admitting dual descriptions
in terms of branes (as in \berggab), where
the tachyon condensation corresponds to some kind of brane
annihilation process.    

It would be interesting to see if these approaches to understanding
closed string tachyons can teach one about string theory at the
Hagedorn transition.  A different approach to this problem was recently 
discussed in \adk.

\medskip
\centerline{\bf{Acknowledgements}}

We would like to thank O. Aharony, P. Aspinwall, M. Berkooz, 
J. Harvey, I. Klebanov, 
E. Martinec, R. Plesser, and S. Shenker  
for discussions.  
We are also grateful to TASI-99 and the 1999 Amsterdam Summer Workshop
on String Theory for hospitality during the course of this work. 
The research of S.K. is supported by NSF grant PHY-95-14797, by
DOE contract DE-AC03-76SF00098, by an A.P. Sloan Foundation Fellowship
and by a DOE OJI Award.
The research of J.K. is supported by the Department of Defense
NDSEG Fellowship Program and by NSF grant PHY-9870115.
The research
of E.S. is supported by the DOE under contract DE-AC03-76SF00515 
and by an
A.P. Sloan Foundation Fellowship.

\listrefs
\end